\documentclass[prl,aps,amssymb,amsmath,superscriptaddress,twocolumn]{revtex4}
\usepackage{amsmath}
\usepackage{amssymb}
\usepackage{epsfig}
\usepackage{amscd}
\usepackage{graphicx,psfrag,xspace}
\usepackage{float}
\usepackage[normalem]{ulem}

\newcommand{\ind}{1\hspace{-2.3mm}{1}}

\begin{document}

\title{Collision number statistics for transport processes}
\author{A. Zoia}
\email{andrea.zoia@cea.fr}
\affiliation{CEA/Saclay, DEN/DANS/DM2S/SERMA/LTSD, 91191 Gif-sur-Yvette, France}
\author{E. Dumonteil}
\affiliation{CEA/Saclay, DEN/DANS/DM2S/SERMA/LTSD, 91191 Gif-sur-Yvette, France}
\author{A. Mazzolo}
\affiliation{CEA/Saclay, DEN/DANS/DM2S/SERMA/LTSD, 91191 Gif-sur-Yvette, France}

\begin{abstract}
Many physical observables can be represented as a particle spending some random time within a given domain. For a broad class of transport-dominated processes, we detail how it is possible to express the moments of the number of particle collisions in an arbitrary volume in terms of repeated convolutions of the ensemble equilibrium distribution. This approach is shown to generalize the celebrated Kac formula for the moments of residence times, which is recovered in the diffusion limit. Some practical applications are illustrated for bounded, unbounded and absorbing domains.
\end{abstract}
\maketitle

The evolution of complex physical systems can often be described in terms of particles undergoing random displacements. Randomness may reflect the intrinsic stochastic nature of the transport process, or result from the uncertainty affecting the interactions between the travelling particle and the surrounding environment~\cite{hughes, weiss}. In this context, assessing the statistics of the residence time $t_{\cal V}$ spent by the random walker inside a given domain ${\cal V}$ plays a key role in many practical problems, encompassing areas as diverse as research strategies, market evolution, percolation through porous media, and DNA translocation through nanopores, to name only a few~\cite{schlesinger, bouchaud_desorder, bouchaud_finance, zoia}. This has motivated a large number of theoretical investigations over the last decade, covering both homogeneous and heterogeneous, scale-invariant media~\cite{redner, avraham, benichou}. In the former case, the dynamics of the walker is typically modelled by regular Brownian motion~\cite{feller}, whereas in the latter one resorts to anomalous diffusion~\cite{klafter}. For Brownian motion, in particular, a seminal work developed by Kac~\cite{kac} has allowed all the moments of the residence times to be evaluated by resorting to convolution integrals over the ensemble equilibrium distribution of the walkers, for arbitrary boundary conditions on ${\cal V}$~\cite{berezhkovskii}. When the particle is lost upon touching the boundary $\partial {\cal V}$ of ${\cal V}$, the residence time is then usually called first-passage time~\cite{redner}, and this quantity has been extensively studied for both Brownian motion and anomalous diffusion~\cite{condamin, benichou, avraham}.

\begin{figure}[b]
   \centerline{\epsfclipon \epsfxsize=6.0cm
\epsfbox{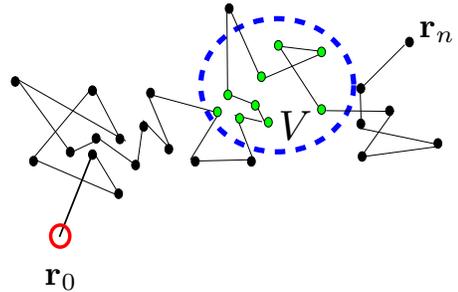} }
   \caption{A random walk starting from ${\mathbf r}_0$ and performing a limited number $n_{\cal V}$ of collisions in a region ${\cal V}$ (with transparent boundaries), before being absorbed at ${\mathbf r}_n$.}
   \label{fig1}
\end{figure}

However, in many realistic situations, stochastic transport is dominated by finite-speed effects, so that the diffusion limit is possibly not attained, and the dynamics of the walker is better described in terms of the Boltzmann equation, rather than the (anomalous) Fokker-Planck equation~\cite{cercignani, case}. Examples are widespread, and arise in, e.g., gas dynamics, neutronics and radiative transfer, electronics, and biology~\cite{cercignani_gas, wigner, jacoboni_book, lecaer}. In all such systems, the stochastic path can be thought of as a series of straight-line flights, separated by random collisions, as in Fig.~\ref{fig1}. A natural variable for describing the walker evolution is therefore the number of collisions $n_{\cal V}$ within the observed volume. Application of the diffusion approximation to the characterization of the residence times distribution, which amounts to assuming a large number of collisions in ${\cal V}$, might lead to inaccurate results~\cite{blanco}. In the present Letter, we address the issue of generalizing Kac approach to random walkers obeying the Boltzmann equation. We derive an explicit formula relating the moments of $n_{\cal V}$ to the equilibrium distribution of the walkers, for both scattering and absorbing media. Knowledge of higher order moments allows on one hand estimating the uncertainty on the average residence time, and on the other hand reconstructing the full distribution of the collision number. We show that when ${\cal V}$ is large as compared to the typical size of a flight, so that the diffusion limit is reached, Kac formula is recovered.

{\em Methodology.} Consider the random walk of a particle starting from a point-source located in ${\mathbf r}_0$. At each collision, the particle can be either scattered, with probability $p$, or absorbed (in which case the trajectory terminates). For the sake of simplicity, we assume that scattering is isotropic. We denote by ${\mathbf r}$ the position of the walker entering a collision, as customary. Let $\pi({\mathbf r},{\mathbf r'})$ be the probability density of performing a displacement from ${\mathbf r'}$ to ${\mathbf r}$, between any two collisions. Define the transport operator $\pi[f]({\mathbf r})$
\begin{equation}
\pi[ f] ({\mathbf r})= \int_{\cal V} \pi({\mathbf r},{\mathbf r'}) f({\mathbf r'}) d{\mathbf r'},
\label{pi_operator}
\end{equation}
over a $d-$dimensional volume ${\cal V}$. We can then express the propagator $\Psi({\mathbf r},n|{\mathbf r}_0)$, i.e., the probability density of finding a particle in ${\mathbf r}$ at the $n-$th collision (starting from ${\mathbf r}_0$), as
\begin{equation}
\Psi({\mathbf r},n|{\mathbf r}_0) = p^{n-1} \pi^n [\delta] ({\mathbf r},{\mathbf r}_0),
\end{equation}
where $\pi^n[f]({\mathbf r})$ is the $n$-th iterated operator
\begin{equation}
\pi^n[ f] ({\mathbf r})= \int_{\cal V}\int_{\cal V}\pi({\mathbf r},{\mathbf r}_{n}) ... \pi({\mathbf r}_2,{\mathbf r}_1)f({\mathbf r}_1) d{\mathbf r}_1 ... d{\mathbf r}_{n},
\label{pi_operator_iterated}
\end{equation}
and $\delta = \delta({\mathbf r}-{\mathbf r}_0)$ is a short-hand for the initial point-source condition.

\begin{figure}[t]
   \centerline{ \epsfclipon \epsfxsize=9.0cm
\epsfbox{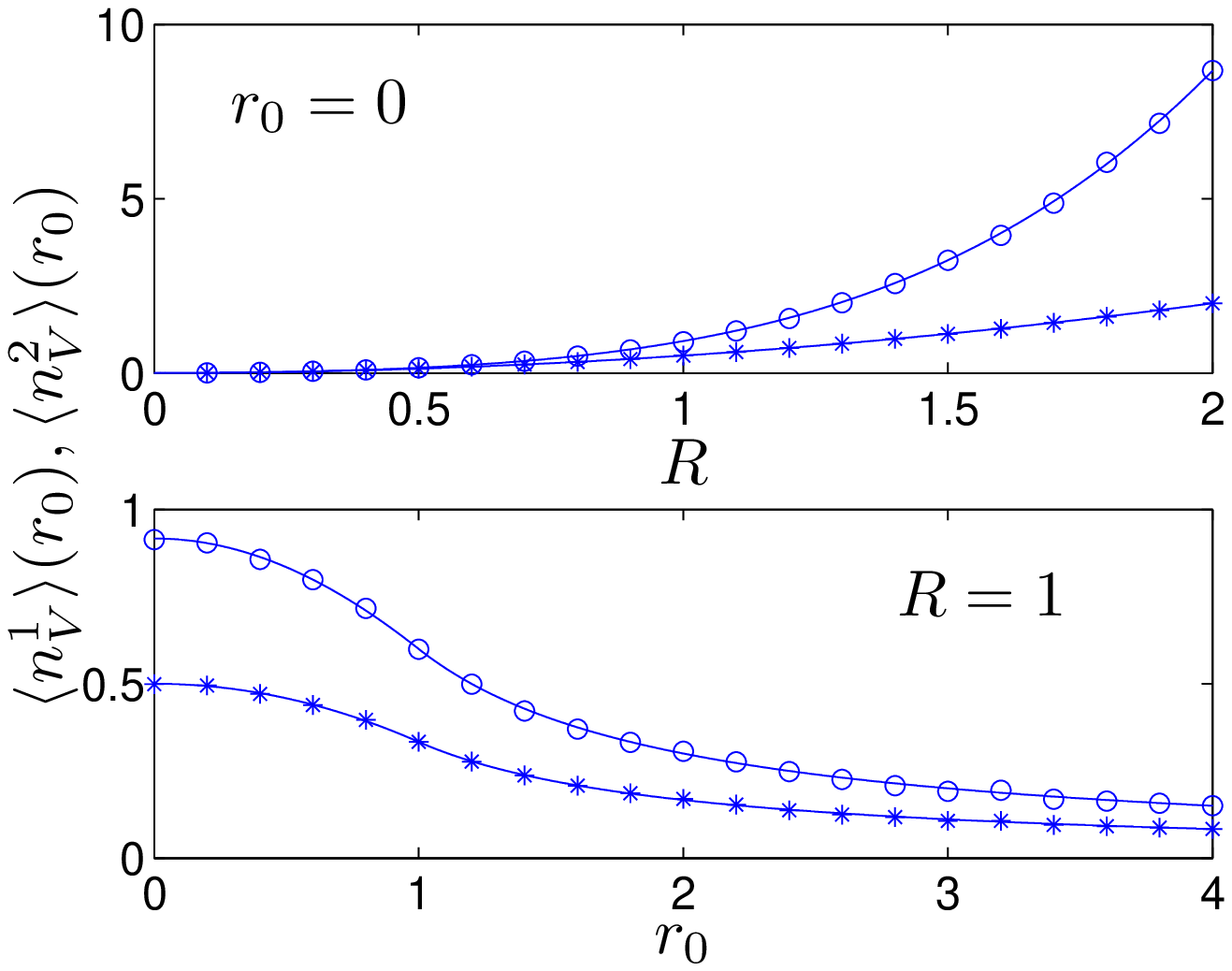} }
\caption{Moments $\langle n_{\cal V}^1 \rangle({\mathbf r}_0)$ (stars) and $\langle n_{\cal V}^2 \rangle({\mathbf r}_0)$ (circles) for $3d$ $\alpha=2$ gamma flights with scattering probability $p=1$ in a volume ${\cal V}$ with transparent boundaries. Theoretical predictions: solid lines; Monte Carlo simulations: symbols.}
   \label{fig2}
\end{figure}

The probability of performing $n_{\cal V}$ collisions in the volume ${\cal V}$ is related to the propagator by
\begin{equation}
{\cal P}(n_{{\cal V}}|{\mathbf r}_0)=\int_{{\cal V}}d {\mathbf r} \Psi({\mathbf r},n_{{\cal V}}|{\mathbf r}_0)-\int_{{\cal V}}d {\mathbf r} \Psi({\mathbf r},n_{{\cal V}}+1|{\mathbf r}_0).
\label{pnv}
\end{equation}
The moments
\begin{equation}
\langle n_{\cal V}^m \rangle({\mathbf r}_0) = \sum_{n_{{\cal V}}=1}^{+\infty} n_{{\cal V}}^m {\cal P}(n_{{\cal V}}|{\mathbf r}_0)
\label{n_poisson}
\end{equation}
depend on the boundary conditions on $\partial {\cal V}$, which affect the functional form of the propagator. The absence of boundary conditions corresponds to defining a fictitious (`transparent') volume ${\cal V}$, where particles can indefinitely cross $\partial {\cal V}$ back and forth. On the contrary, the use of leakage boundary conditions leads to the formulation of first-passage problems~\cite{redner}.

We introduce now the collision density $\Psi({\mathbf r}|{\mathbf r}_0)$
\begin{equation}
\Psi({\mathbf r}|{\mathbf r}_0)=\lim_{N \to \infty} \sum_{n=1}^N \Psi({\mathbf r},n|{\mathbf r}_0),
\label{collision_density}
\end{equation}
which intuitively represents the equilibrium (stationary) particle distribution~\cite{case}. From the definition of $\Psi({\mathbf r}|{\mathbf r}_0)$, it follows immediately that
\begin{equation}
\langle n^1_{\cal V} \rangle({\mathbf r}_0) = \int_{{\cal V}}d {\mathbf r} \Psi({\mathbf r}|{\mathbf r}_0),
\label{average_n}
\end{equation}
i.e., the integral of the collision density over a volume ${\cal V}$ gives the mean number of collisions within that domain, hence the name given to $\Psi({\mathbf r}|{\mathbf r}_0)$.

\begin{figure}[t]
   \centerline{ \epsfclipon \epsfxsize=9.0cm
\epsfbox{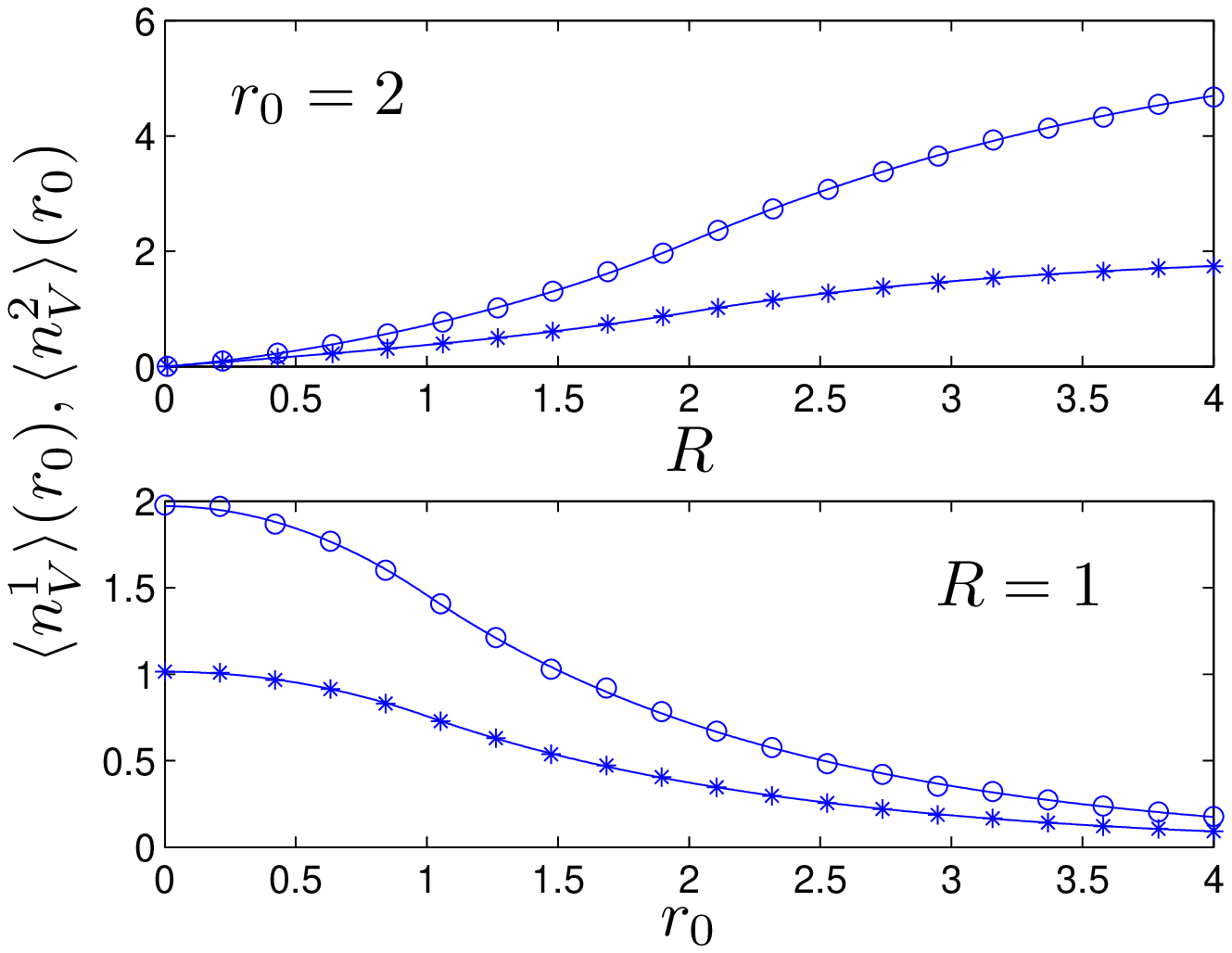} }
   \caption{Moments $\langle n_{\cal V}^1 \rangle({\mathbf r}_0)$ (stars) and $\langle n_{\cal V}^2 \rangle({\mathbf r}_0)$ (circles) for $1d$ exponential flights with scattering probability $p=0.5$ in a volume ${\cal V}$ with transparent boundaries. Theoretical predictions: solid lines; Monte Carlo simulations: symbols.}
   \label{fig3}
\end{figure}

Higher order moments of $n_{\cal V}$ can be obtained as follows. Define the operator
\begin{equation}
\Psi[f]({\mathbf r}) = \int_{\cal V} \Psi({\mathbf r}|{\mathbf r'}) f({\mathbf r'}) d{\mathbf r'}.
\end{equation}
By making use of the Neumann series
\begin{equation}
\sum_{n=1}^{\infty} p^{n-1}\pi^n [f]({\mathbf r})= \frac{\pi}{1-p\pi}[f]({\mathbf r}),
\end{equation}
we have then
\begin{equation}
\Psi[f] ({\mathbf r})= \frac{\pi}{1-p\pi}[f] ({\mathbf r}),
\end{equation}
and in particular $\Psi({\mathbf r}|{\mathbf r}_0) = \Psi[\delta]({\mathbf r},{\mathbf r}_0)$. Now, combining Eqs.~\eqref{n_poisson} and ~\eqref{pnv}, we get
\begin{equation}
\langle n_{\cal V}^m \rangle({\mathbf r}_0) = \frac{1}{p}\int_{\cal V} d{\mathbf r}\operatorname{Li}_{-m}(p\pi) (1-p\pi)[\delta] ({\mathbf r},{\mathbf r}_0),
\label{moment_2}
\end{equation}
where $\operatorname{Li}_{s}(x)=\sum_{k=1}^{\infty}x^k / k^s$ is the polylogarithm function~\cite{erdelyi}. When $m$ is a non-negative integer, the polylogarithm is a rational function, namely,
\begin{equation}
\operatorname{Li}_{-m}(x)=\sum_{k=0}^{m} k! s_{m+1,k+1} \left( \frac{x}{1-x}\right)^{k+1} ,
\end{equation}
where the coefficients
\begin{equation}
s_{m,k}=\frac{1}{k!} \sum_{i=0}^{k} (-1)^i \binom {k} {i} \left( k-i\right)^m
\end{equation}
are the Stirling numbers of second kind~\cite{erdelyi}. Thanks to the recurrence properties of the Stirling numbers, Eq.~\eqref{moment_2} gives
\begin{equation}
\langle n_{\cal V}^m \rangle({\mathbf r}_0) = \frac{1}{p} \int_{\cal V} d{\mathbf r} \sum_{k=1}^{m} k! s_{m,k} p^k \Psi^{k} [\delta]({\mathbf r},{\mathbf r}_0).
\label{moment_4}
\end{equation}
We introduce then the repeated Kac integrals
\begin{equation}
{\cal C}_k({\mathbf r}_0) = k! \int_{{\cal V}}d {\mathbf r}_k ... \int_{{\cal V}} d {\mathbf r}_1 \Psi({\mathbf r}_k|{\mathbf r}_{k-1})...\Psi({\mathbf r}_1|{\mathbf r}_0),
\label{kac_integrals}
\end{equation}
which are defined as $k$-fold convolutions of the collision density $\Psi({\mathbf r}|{\mathbf r}_0)$ with itself~\cite{kac, berezhkovskii}. It follows the equivalence
\begin{equation}
{\cal C}_k({\mathbf r}_0) = k! \int_{\cal V} d{\mathbf r} \Psi^{k} [\delta]({\mathbf r},{\mathbf r}_0),
\label{kac_integrals_operator}
\end{equation}
with ${\cal C}_{k+1}({\mathbf r}_0)=k\Psi[{\cal C}_k]({\mathbf r}_0)$ and ${\cal C}_0({\mathbf r}_0)=\ind_{{\mathbf r}_0 \in {\cal V}}$, $\ind$ being the characteristic function. The convergence of the integrals ${\cal C}_k({\mathbf r}_0)$ depends on the features of the underlying stochastic process as well as on boundary conditions. For instance, the persistence property of walks in $d \le 2$~\cite{feller} implies diverging ${\cal C}_k({\mathbf r}_0)$ for transparent ${\cal V}$ in absence of absorption~\cite{berezhkovskii}. We finally obtain the central result of this Letter, i.e., the explicit formula for the moments of the collision number in ${\cal V}$
\begin{equation}
\langle n_{\cal V}^m \rangle({\mathbf r}_0) = \frac{1}{p} \sum_{k=1}^{m} s_{m,k} p^k {\cal C}_k({\mathbf r}_0).
\label{n_moments}
\end{equation}

\begin{figure}[t]
   \centerline{ \epsfclipon \epsfxsize=9.0cm
\epsfbox{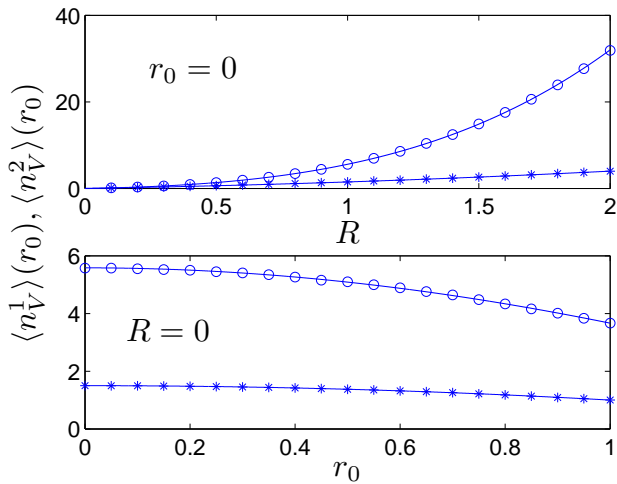} }
   \caption{Moments $\langle n_{\cal V}^1 \rangle({\mathbf r}_0)$ (stars) and $\langle n_{\cal V}^2 \rangle({\mathbf r}_0)$ (circles) for $1d$ exponential flights with scattering probability $p=1$ in a volume ${\cal V}$ with leakage boundaries. Theoretical predictions: solid lines; Monte Carlo simulations: symbols.}
   \label{fig4}
\end{figure}

Thanks to linearity, Eq.~\eqref{n_moments} allows expressing $\langle n_{\cal V}^m \rangle({\mathbf r}_0)$ as a combination of $m$ Kac integrals, $k=1,...,m$, each given from Eq.~\eqref{kac_integrals}. In particular, for $m=1$ we recover Eq.~\eqref{average_n}, since $s_{1,1}=1$. Furthermore, for $m=2$, $s_{2,1}=s_{2,2}=1$, so that
\begin{equation}
\langle n_{\cal V}^2 \rangle = p 2!\int_{{\cal V}} d {\mathbf r}_2 \int_{{\cal V}} d {\mathbf r}_1 \Psi({\mathbf r}_2|{\mathbf r}_1)\Psi({\mathbf r}_1,{\mathbf r}_0) +\langle n_{\cal V}^1 \rangle.
\end{equation}
Let $G(z|{\mathbf r}_0)$ be the moment generating function of ${\cal P}(n_{{\cal V}}|{\mathbf r}_0)$~\cite{feller}. By definition we have the moment expansion
\begin{equation}
G(z|{\mathbf r}_0)= \sum_{m=0}^{\infty} \langle n_{\cal V}^m \rangle({\mathbf r}_0) \frac{z^m}{m!}.
\end{equation}
The small-$z$ expansion of $G(z|{\mathbf r}_0)$ reads then $G(z|{\mathbf r}_0) \simeq 1 + \langle n_{\cal V}^1 \rangle({\mathbf r}_0) z$, which for the Tauberian thorems corresponds to the large-$n_{\cal V}$ behavior. It follows the exponential tail
\begin{equation}
{\cal P}(n_{{\cal V}}|{\mathbf r}_0) \simeq e^{-n_{{\cal V}}/\langle n_{\cal V}^1 \rangle({\mathbf r}_0)},
\label{eq_p_n_tail}
\end{equation}
provided that ${\cal C}_1({\mathbf r}_0)$ is finite.

{\em Diffusion limit and Kac formula.} Suppose for the sake of simplicity that the walker moves at constant speed $v$, so that the time spent between any two collisions is $t_i=|{\mathbf r}_i-{\mathbf r}_{i-1}|/v$. For isotropic walks, $\pi({\mathbf r},{\mathbf r}')=\pi(\ell=|{\mathbf r}-{\mathbf r}'|)$ thanks to the spherical symmetry. Then, flight times are identically distributed, and obey $t_i \sim w(t_i)$, where $w(t_i)=\Omega_d\int \ell^{d-1}\pi(\ell)\delta(t_i-\ell/v)d\ell$, $\Omega_d$ being the surface of the unit sphere. The diffusion limit of the transport process described above is obtained by letting the typical flight length $\sigma$ (the standard deviation of jump sizes, as customary) and the average inter-collision time $\tau = \langle t_i \rangle$ shrink to zero in such a way that the ratio $D = \sigma^2 / \tau$ converges to a constant, namely, the diffusion coefficient. When $\tau$ and $\sigma$ vanish, the collision number in ${\cal V}$ diverges, whereas the quantity
\begin{equation}
t_{{\cal V}} = \sum_{i=1}^{n_{{\cal V}}} t_i
\label{tn}
\end{equation}
converges to the residence time in the volume. Actually, $t_{{\cal V}}$ should take into account also additional terms due to boundary conditions. However, as $\tau \to 0$ and $n_{{\cal V}} \to \infty$, the trajectory will almost surely have a turning point touching the boundary, so that corrections to Eq.~\eqref{tn} can be safely neglected. Let now ${\cal Q}(t_{\cal V}|{\mathbf r}_0)$ be the distribution of the residence times. Under the previous assumptions, in the Laplace space we have ${\cal Q}(s|{\mathbf r}_0) =\int \exp(-st_{{\cal V}}){\cal Q}(t_{{\cal V}}|{\mathbf r}_0)dt_{{\cal V}} =w(s)^{n_{{\cal V}}}$. Any arbitrary $w(s)$ with finite $\tau$ has an expansion $w(s) \simeq 1 - s\tau$ when $\tau \to 0$. Then we have ${\cal Q}(s|{\mathbf r}_0) \simeq e^{-n_{\cal V} s\tau}$, which implies ${\cal Q}(t_{\cal V}|{\mathbf r}_0) \simeq \delta(t_{\cal V}-n_{\cal V}\tau)$ for small $\tau$. It follows that
\begin{equation}
\langle t^m_{{\cal V}} \rangle ({\mathbf r}_0) \simeq \tau^m \langle n_{\cal V}^m \rangle ({\mathbf r}_0),
\end{equation}
with $\langle n_{\cal V}^m \rangle ({\mathbf r}_0)$ given by Eq.~\eqref{n_moments}. If we rescale the space variable ${\mathbf r}$ by $\sigma$, each of the terms of the sum in Eq.~\eqref{n_moments} carries a contribution $\sigma^{-2k}$. In the diffusion limit, we therefore recover the celebrated Kac formula~\cite{kac, berezhkovskii} for the moments of the residence times of Brownian motion
\begin{equation}
\langle t^m_{{\cal V}} \rangle ({\mathbf r}_0) \simeq \frac{{\cal C}_m ({\mathbf r}_0)}{D^m},
\end{equation}
because all other terms in the sum vanish when $\tau \to 0$ and $\sigma \to 0$, and $s_{m,m}=1$. Moreover, we have the recursion property $D \langle t^{m+1}_{{\cal V}} \rangle ({\mathbf r}_0)=m\Psi[\langle t^{m}_{{\cal V}} \rangle]({\mathbf r}_0)$, starting from $\langle t^{0}_{{\cal V}} \rangle ({\mathbf r}_0)=\ind_{{\mathbf r}_0 \in {\cal V}}$. As for the residence time distribution, from Eq.~\eqref{eq_p_n_tail} at long times we have ${\cal Q}(t_{{\cal V}}|{\mathbf r}_0) \simeq \exp(-Dt_{{\cal V}}/{\cal C}_1({\mathbf r}_0))$.

{\em Discussion and perspectives.} For simple geometries and $\Psi({\mathbf r}|{\mathbf r}_0)$, formula~\eqref{n_moments} is amenable to analytical solutions. Here we illustrate some significant cases, ${\cal V}$ being a $d$-sphere of radius $R$ centered in $\bf 0$. Isotropic gamma flights with kernel $\pi(\ell)=\ell^{\alpha-d}\exp(-\ell)/\Omega_d \Gamma(\alpha)$, $\alpha>0$, are a widespread transport process and describe, among others, search strategies in biology~\cite{gamma_biology}. When $\alpha=2$, the $3d$ scattering collision density assumes a simple form,
\begin{equation}
\Psi({\mathbf r}|{\mathbf r}_0)=\frac{1}{4\pi |{\mathbf r}-{\mathbf r}_0|}.
\end{equation}
It follows
\begin{equation}
\langle n^1_{{\cal V}} \rangle ({\mathbf r}_0) = \begin{cases} \frac{3R^2-r_0^2}{6} & r_0<R  \\ \frac{R^3}{3 r_0} & r_0\ge R \end{cases}
\end{equation}
and
\begin{equation}
\langle n^2_{{\cal V}} \rangle ({\mathbf r}_0) = \begin{cases} \frac{25R^4-10R^2r_0^2+r_0^4}{60}+\langle n^1_{{\cal V}} \rangle ({\mathbf r}_0)& r_0<R  \\ \frac{4}{15}\frac{R^5}{r_0}+\langle n^1_{{\cal V}} \rangle ({\mathbf r}_0) & r_0\ge R \end{cases}
\end{equation}
Comparisons with Monte Carlo simulations with $10^5$ particles are shown in Fig.~\ref{fig2}. Exponential flights ($\alpha=1$) arise when the scattering centers are uniform, so that inter-collision distances obey a Poisson distribution. Such process is crucial for understanding, e.g., radiation propagation~\cite{cercignani, case}. The transport kernel reads $\pi(\ell)=\ell^{1-d}\exp(-\ell)/\Omega_d$. In $1d$ systems, the collision density for absorbing ${\cal V}$ and transparent boundaries reads
\begin{equation}
\Psi({\mathbf r}|{\mathbf r}_0)=\frac{e^{-\sqrt{1-p}|{\mathbf r}-{\mathbf r}_0|}}{2\sqrt{1-p}},
\end{equation}
whereas for scattering ${\cal V}$ and leakage boundaries
\begin{equation}
\Psi({\mathbf r}|{\mathbf r}_0)=\frac{r_e-|{\mathbf r}-{\mathbf r}_0|}{2},
\end{equation}
with $r_e=R+1$~\cite{zoia_dumonteil_mazzolo}. The moments $\langle n^m_{{\cal V}} \rangle ({\mathbf r}_0)$ are again easily obtained from Eq.~\eqref{n_moments}. The resulting formulas are rather cumbersome and will not be presented here. Comparisons of exact results and Monte Carlo simulations with $10^5$ particles are shown in Figs.~\ref{fig3} and~\ref{fig4}.

The integrals appearing in Eq.~\eqref{n_moments} can be carried out numerically for arbitrary complex volumes and $\Psi({\mathbf r}|{\mathbf r}_0)$. Moreover, the isotropy hypothesis can be possibly relaxed by replacing ${\mathbf r}$ with a state variable ${\mathbf y}=\left\lbrace {\mathbf r}, {\mathbf \theta}\right\rbrace$ accounting for the scattering angle. In this respect, the approach presented in this Letter is general, and applies to a broad class of transport processes and boundary conditions. Finally, we observe that Eq.~\eqref{n_moments} could be used as an estimator to infer the properties of the underlying stochastic path, which is often not accessible, on the basis of the collision statistics in a measure volume.


\end{document}